\pdfoutput=1
\documentclass[usenatbib]{mn2e}
\usepackage[T1]{fontenc}
\usepackage[latin9]{inputenc}
\usepackage{bm}
\usepackage{amsbsy}
\usepackage{amstext}
\usepackage{graphicx}
\usepackage{esint}
\newcommand{\apj}{ApJ}
\newcommand{\araa}{Ann. Rev. Astron and Astroph.}
\newcommand{\pre}{Phys. Rev. E}
\newcommand{\apjl}{ApJL}

\newcommand{\aap}{A~\&~A}
\newcommand{\mnras}{MNRAS}
\newcommand{\na}{New Astr.}
\newcommand{\solphys}{Sol.Phys.}
\newcommand{\physrep}{Phys. Rep.}

\title[Dependence of magnetic cycle parameters on period]{Dependence of magnetic cycle parameters on period of rotation in nonlinear solar-type
dynamos}

\author[V.V.~Pipin]
{V.V. Pipin \thanks{email: pip@iszf.irk.ru}\\
{Institute of Solar-Terrestrial Physics, Russian Academy of
Sciences, Irkutsk, 664033, Russia}\\ }

\begin{document}
\label{firstpage} \maketitle

\providecommand{\tabularnewline}{\\}

\begin{abstract}
Parameters of magnetic activity on the solar type stars
depend on the properties of the dynamo processes operating in 
stellar convection zones. We apply nonlinear mean-field axisymmetric
$\alpha^2\Omega$ dynamo models to calculate of the magnetic cycle parameters, such
as the dynamo cycle period, the total magnetic flux and the
Poynting magnetic energy flux on the surface of solar analogs with the rotation
periods from 15 to 30 days. The  models take into account the
principal nonlinear mechanisms of the large-scale
dynamo, such as the magnetic helicity conservation, magnetic buoyancy,
and effects of
magnetic forces on the angular momentum balance inside the convection
zones. Also, we consider two types of the dynamo models. The
distributed (D-type)
models employ the standard $\alpha$- effect distributed on the whole
convection zone. The ``boundary'' (B-type) models employ the non-local $\alpha$-
effect, which is confined to the boundaries of the convection zone.
Both the D- and B-type models show that the dynamo-generated magnetic
flux increases with the increase of the stellar rotation
rate. {It is found that for the considered range of the rotational
periods} the magnetic helicity conservation is the most significant effect
for the nonlinear  quenching of the dynamo. This quenching is
more efficient in the B-type than in the D-type dynamo models.
The D-type dynamo reproduces the observed dependence of the cycle
period on the rotation rate for the Sun analogs. For the solar analog 
rotating with a period of 15 days we find nonlinear dynamo regimes with multiply cycles.
\end{abstract}

\section{Introduction}

Currently, there is an extensive database of observations of magnetic activity
on the main sequence stars (see, e.g., reviews by \citealt{2009ARAA_donat,lrsp-2012-1}).
Cool stars with outer convective envelope are of particular
interests because they are Sun-like. It is believed that the magnetic
activity on the solar-like stars results from large-scale
dynamo processes driven by turbulent convection and rotation
\citep{2005PhR...417....1B}. 
Observations (e.g., \citep{2007ApJ_bohm,2009ARAA_donat,2010NewAstkazova,2011IAUS_saar,2013GeAe_kazova,2014MNRAS_mars,vid14MNR}),
as well as the 2D mean-field models of the angular momentum balance \citep{1989drsc.book.....R,1999A&A...344..911K,2013IAUS..294..399K}
and the 3D numerical simulations \citep{2009AnRFM..41..317M,2011ApJ...740...12H,2013ApJ...779..176G,2014A&A...570A..43K,raey})
show that parameters of the differential rotation and convection,
e.g., the typical size and turnover time of convective flows,
depend on the general  stellar parameters, such as  mass, 
age , the spectral class and the rotation rate. The mass of a star
and its Rossby number, which is the ratio of the period of rotation
and a typical turnover time of convection, are likely the most important
parameters governing the stellar dynamo \citep{2009ARAA_donat,2013AN....334...48M}.
The diagram 3 in the paper by \citet{2009ARAA_donat} shows an increase
of the magnetic activity with decrease of the Rossby number
and the mass of a star. These parameters determine the topology of
the large-scale magnetic field, as well. It is found that the axisymmetric
solar-type dynamo can operates in stars with mass about $1M_{\odot}$,
and with periods of rotation longer than 10
days. {Observations, also
show evidences that solar analogs with period of rotation smaller than 10 days, may have} the substantial
non-axisymmetric components of the large-scale magnetic field \citep{2009ARAA_donat,2014IAUS..302..110F}.

Interpretation of the stellar magnetic activity is rather complicated
because of  nonlinear dynamo effects that needs to be taken into account 
\citep{soon93,1999ApJ_sa_br,2007ApJ_bohm,2011IAUS_saar,2015MNRAS.446L..51B}.
Moreover, even for the solar dynamo processes many details are poorly known.
\citep{chrev05}. In particular, the origin of the large-scale poloidal
magnetic field of the Sun (the component of the field which lies in meridional
planes) is not well understood. \citet{P55} suggested
that convective vorticies acting on the rising parts of large-scale
toroidal magnetic field produce a ensemble of the small-scale helical
magnetic loops with a North-South field component.
Thus,  the large-scale toroidal field is partially
transformed into the poloidal magnetic field. It is the so-called
$\alpha$- effect. 
In this picture, the
drift of the sunspot formation latitudinal zone (``the butterfly diagrams'') and the reversal of the polar
magnetic field are explained by magnetic diffusion  of the dynamo-wave
travelling  in the convection zone along the isolines of the angular velocity
(\citealp{yosh1975,b05,2006AN....327..884K,2013ASPC..479..395K}).

\citet{1961ApJ...133..572B} suggested an alternative mechanism in which
the poloidal magnetic field originated from buoyant
toroidal field loops which turned by the Coriolis force when they rise
to the surface. This effect can
be considered as a non-local $\alpha$ - effect \citep{bs02,2007NJPh....9..305B,2014arXiv1412.0997B}.
 This type of dynamo (sometimes called  Babcock-Leighton dynamo)
 operates near the boundaries of the convection zone. In this dynamo mechanism the strong
 toroidal field is concentrated in the solar tachocline and the poloidal field is
 generated near the surface.
In the dynamo model with  the non-local $\alpha$ - effect, the solar
type time-latitude butterfly diagrams 
for the toroidal field can be reproduced with (\citealt{choud95,dc99}) and
without meridional circulation \citep{2011AstL...37..286K}. The distribution of the meridional circulation in the
Sun is still not well established. Helioseismology measurements have
deduced a double-cell structure
of the meridional circulation \citep{Zhao13m,2014ApJ...784..145K},
while the 3D MHD simulations suggest an even more complicated
structure (e.g., \citealp{2013ApJ...779..176G,2014A&A...570A..43K,raey}) 

Recently, \citet{2010A&A...510A..33B} and \citep{2014ApJ...791...59K}
discussed results of simulations for the kinematic dynamo models, which
take into account the non-local $\alpha$- effect and the meridional
circulation for the solar-type stars (1M$_{\odot}$) for the range
of the rotational period from 1 to 30 days. These models qualitatively
reproduce the increase of the magnetic activity with the increase of
the rotation rate. However, the models fail to explain the decrease of the
dynamo period with the increase of the rotation rate (and also the magnitude
of the generated magnetic fields). \citet{2010A&A...510A..33B} found
that the issue can be solved by assuming certain multiply-cell pattern
of the meridional circulation.

The reverse correlation between the magnetic cycle amplitude and the
cycle period is a common feature of the solar \citep{vetal86} and
stellar \citep{soon94} magnetic cycles. It is fulfilled for the dynamo
model with the distributed $\alpha$- effect \citep{pk11,pipea2012AA}.
A goal  of this paper is to examine this relation for a
range of the rotation rates faster than the solar rotation.

The study explores the rotational periods from 15 to 30 days. We
assume that the initial structure of the differential rotation corresponds
to the current results from helioseismology \citep{Howe2011JPh}.
Results of \citet{2014ApJ...791...59K} confirm this assumption, which is also supported
by observations \citep{2011IAUS_saar}.

The paper studies a set of  nonlinear dynamo models which take
into account the magnetic feedback on the angular momentum balance,
the magnetic helicity conservation and the magnetic buoyancy
effect.  Two type of the dynamo models will be discussed:
1) the D-type model, with the $\alpha$-effect in the bulk of the
convection zone; 2) the B-type model with the non-local $\alpha$-
effect.
 Confronting
two approaches allows us to explore how distribution of the turbulent
sources of the dynamo impact the properties of the magnetic cycles.
Also, it is interesting to confront the nonlinear saturation in the
different types of the mean-field dynamos. Our results are preliminary
because the  meridional circulation is not included  in the study.

\section{Basic equations}

\subsection{Dynamo model\label{DM}}

Following \citet{KR80}, we explore the evolution of the induction
vector of the mean field, $\overline{\mathbf{B}}$, in the highly conductive
turbulent media with the mean flow velocity $\overline{\mathbf{U}}$ and the
mean electromotive force, $\boldsymbol{\mathcal{E}}=\overline{\mathbf{u}\times\mathbf{b}}$
(hereafter MEMF), where $\mathbf{u}$ and $\mathbf{b}$ are fluctuations
of the flow and magnetic field: 
\begin{equation}
\frac{\partial\overline{\mathbf{B}}}{\partial t}=\boldsymbol{\nabla}\times\left(\boldsymbol{\mathcal{E}}+\overline{\mathbf{U}}\times\overline{\mathbf{B}}\right)\text{.}\label{eq:dyn}
\end{equation}
We employ a decomposition of the axisymmetric field into a sum of
the azimuthal(toroidal) and poloidal (meridional plane) components: 
\[
\overline{\mathbf{B}}=\mathbf{e}_{\phi}B+\nabla\times\frac{A\mathbf{e}_{\phi}}{r\sin\theta},
\]
where $\mathbf{e}_{\phi}$ is the unit vector in the azimuthal direction,
$\theta$ is the polar angle and $A\mathbf{e}_{\phi}$ is the vector-potential
of the large-scale poloidal magnetic field. The mean electromotive
force, $\boldsymbol{\mathcal{E}}$, is expressed as follows: 
\begin{equation}
\mathcal{E}_{i}=\left(\alpha_{ij}+\gamma_{ij}\right)\overline{B}_{j}-\eta_{ijk}\nabla_{j}\overline{B}_{k}.\label{eq:EMF-1}
\end{equation}
The tensor, $\alpha_{ij}$, models the generation of the magnetic field
by the $\alpha$- effect; the anti-symmetric tensor, $\gamma_{ij}$,
controls pumping of the large-scale magnetic fields in turbulent
media; the tensor, $\eta_{ijk}$, governs the turbulent diffusion and takes into account
the generation of the magnetic fields by the $\Omega\times J$ effect
\citep{rad69}, as well. We take
into account the effect of rotation and magnetic field on the mean-electromotive
force. The technical details can be found in our previous papers \citep{pi08Gafd,pipea2012AA,pip13M,2014ApJ_pipk}.
The expressions for the tensor  coefficients are given in Appendix.

For the B-type models we employ the following expression for the
mean-electromotive force related with the non-local $\alpha$-effect
\citep{2007NJPh....9..305B,2014arXiv1412.0997B}:

\begin{eqnarray}
\mathcal{E}^{(\alpha)}{}_{\phi} & = & C_{S}\cos\theta\psi^{\left(-\right)}\left(r-r_{s}\right)\label{eq:babck}\\
 & \times & \int_{r_{b}}^{r}\frac{\psi^{\left(+\right)}\left(r'-r_{bt}\right)\!B\left(r',\theta\right)\!}{1+B^{2}\left(r',\theta\right)}\!d\!r'+{\displaystyle \frac{\overline{\chi}B\left(r,\theta\right)}{\overline{\rho}\ell^{2}}},\nonumber 
\end{eqnarray}
where $C_{S}$ is a free parameter that controls the strength of the $\alpha$-
effect, the $\psi^{\left(\pm\right)}=\frac{1}{2}\left(1\pm\mathrm{erf}\left(100r\right)\right)$
are the Heaviside-like functions, with $r_{s}=0.95R$,
$r_{bt}=0.725R$ and the
$\overline{\chi}=\overline{\mathbf{a}\cdot\mathbf{b}}$ is helicity density
  of the fluctuating part of magnetic field with the  $\mathbf{a}$ being a fluctuating
  vector-potential.
 { Following results of \citet{2014ApJ...791...59K} we
assume that the strength of the non-local $\alpha$-effect does not
change significantly in the explored interval of the rotation rates.}

In our models the $\alpha$ effect takes into account the kinetic and
magnetic helicities.  
{Conservation of the magnetic helicity results to
  the dynamical quenching of the $\alpha$-effect.} The helicity density
  of the fluctuating part of magnetic field,  $\overline{\chi}$,
is governed by the conservation law \citep{hub-br12}: 
\begin{equation}
\frac{\partial\overline{\chi}^{(tot)}}{\partial t}=-\frac{\overline{\chi}}{R_{m}\tau_{c}}-2\eta\overline{\mathbf{B}}\cdot\mathbf{\overline{J}}-\boldsymbol{\nabla\cdot}\boldsymbol{\boldsymbol{\mathcal{F}}}^{\chi},\label{eq:helcon-1}
\end{equation}
where $\overline{\chi}^{(tot)}=\overline{\chi}+\overline{\mathbf{A}}\cdot\overline{\mathbf{B}}$
is the total magnetic helicity density and the $\boldsymbol{\boldsymbol{\mathcal{F}}}^{\chi}$
is the diffusive flux of the magnetic helicity, $R_{m}$
is the magnetic Reynolds number, and $\eta$ is the microscopic
magnetic diffusivity. 

The distribution of the turbulent parameters, e.g, {such as
the typical convective turn-over time $\tau_{c}$, the mixing length
$\ell$, the RMS convective velocity $u'$, the mean density $\bar\rho$
and its gradient
$\mathbf{\boldsymbol{\Lambda}}^{(\rho)}=\boldsymbol{\nabla}\log\overline{\rho}$
are computed from the mixing-length model of the solar convection zone
of  \cite{stix:02}. In particular, it uses the mixing length
$\ell=\alpha_{{\rm MLT}}\left|\Lambda^{(p)}\right|^{-1}$,
 where $\Lambda{}^{(p)}={\nabla}\log\overline{p}$, 
is the reverse scale of the thermodynamic pressure and $\alpha_{{\rm MLT}}=2$.}
The profile of the turbulent diffusivity is given in form $\eta_{T}=C_{\eta}{\displaystyle \frac{u'^{2}\tau_{c}}{3f_{ov}\left(r\right)}}$,
where $f_{ov}(r)=1+\exp\left(50\left(r_{ov}-r\right)\right)$, $r_{ov}=0.725R_{\odot}$
controls quenching of the turbulent effects near the bottom of the
convection zone. The parameter $C_{\eta}$, (with  the range $0<C_{\eta}<1$) is a free parameter
to control the efficiency of the mixing of the large-scale magnetic
field by the turbulence. It is used to adjust the period
of the dynamo cycle. We use the same model of the convection zone for all the rotational periods.

At the bottom of the convection zone we apply the superconductor
boundary condition both in the case of the D- and B-types dynamos. At the top of the convection zone the poloidal
field is smoothly matched to the external potential field and the
toroidal field is allowed to penetrated to the surface:
\begin{eqnarray}
\left(\delta+\frac{B}{B_{esq}}\right)\frac{\eta_{T}}{r_{e}}B+\left(1-\delta\right)\mathcal{E}_{\theta} & = & 0,\label{eq:tor-vac}
\end{eqnarray}
where $\delta=0.99$, $B_{esq}=10$G. For the $B\gg B_{esq}$ the
condition (\ref{eq:tor-vac}) is close to the vacuum boundary conditions
\citep{1992A&A...256..371M}. The given boundary conditions provide the
non-zero Poynting magnetic energy flux 
$S=-{\displaystyle \frac{1}{4\pi}\left.\mathcal{E}_{\theta}B\right|_{r_{e}}}$
through the top. In the D-type models we  estimate the total
Poynting flux by integrating over the solid angle, 
 $S_M=\displaystyle{2\pi\int_0^{\pi}\sin\theta\overline{S(\theta )  }
   \mathrm{d} \theta}$, where
  $\overline{S(\theta)}$ is the mean Poynting flux in subsurface shear
  layer, for the range of radial distances from 0.9 to 0.99R. For the B-type models we employ the standard vacuum
boundary conditions at the top of the domain. 
\begin{table*}
\protect\caption{Summary of the dynamo models and their parameters
  (see Appendix for details).}
\begin{tabular}{|c|c|c|}
\hline 
MEMF parts & D-type  & B-type \\
\hline
$\eta_{ijk}$ -diffusivity & Anisotropic, Eq.(\ref{eq:diff}) & Eq.(\ref{eq:diff}), $C_{\delta}=0$, $a=0$\\
$\alpha_{ij}$-effect & Distributed Eq.(A2) & Non-local Eq.(\ref{eq:babck})\\
$\gamma_{ij}$ -pumping & Eqs.(\ref{eq:pump}) & Eqs.(\ref{eq:pumpu},\ref{eq:pumpb})\\
free parameters &
$C_{\alpha}=0.04$,$C_{\delta}=\displaystyle{C_{\alpha}\over 3}$, $C_{\eta}=\displaystyle{1\over 15}$, $C_{v}=\displaystyle{1\over 2}$,
a=3 & $C_{\eta}=\displaystyle{1\over 15}$, $C_{S}=4$ ,$C_{v}=4$\\
\hline 
\end{tabular}
\end{table*}

\subsection{Angular momentum balance}

The mean angular momentum balance in the stellar convection zone is established due to the turbulent
stresses $T_{r\phi}$, $T_{\theta\phi}$, the meridional circulation and
the large-scale Lorentz force \citep{1989drsc.book.....R} :
\begin{eqnarray}
\sin\theta\frac{\partial\Omega}{\partial t} & = & \frac{1}{\rho
  r^{4}}\frac{\partial}{\partial r}r^{3}\left(T_{r\phi}
-\frac{B}{4\pi r^{2}}\frac{\partial A}{\partial\mu}\right)\label{eq:angm}\\
 & - & \frac{1}{\rho r^{2}\sin\theta}\frac{\partial}{\partial\mu}\left(\sin^{2}\theta T_{\theta\phi}-\frac{B}{4\pi r\sin\theta}\frac{\partial A}{\partial r}\right)\nonumber 
\end{eqnarray}
where the turbulent stresses $T_{r\phi}$ and $T_{\theta\phi}$ take
into account the turbulent viscosity and generation of the large-scale 
shear due to the $\Lambda$- effect. 

{Following to  \citet{1975JFM....67..417M}, \citet{tob1996} and
  \citet{cov00}, it is assumed
that in the absence of the large-scale magnetic fields the solution of
the Eq.(\ref{eq:angm})
is time-independent  and it describes the steady angular velocity
profile. We apply the same initial {\it dimesionless} profile of the
angular velocity, which was
taken from \citet{Howe2011JPh}, to all rotational periods studied in
the paper. This profile is used as the reference
state. We subtract the reference state from the
Eq.(\ref{eq:angm}) to find the equation for perturbations
of the angular velocity profile from the reference state due to effects
of the large- and small-scale Lorentz forces}: 
\begin{eqnarray}
\sin\theta\frac{\partial\delta\Omega}{\partial t} & = & \frac{1}{\rho r^{4}}\frac{\partial}{\partial r}r^{3}\bigg(T_{r\phi}\left(\Omega_{\odot}+\delta\Omega,\Omega^{*},\beta\right)\\
 & - & T_{r\phi}\left(\Omega_{\odot},\Omega^{\star},\beta=0\right)-\frac{B}{4\pi r^{2}}\frac{\partial A}{\partial\mu}\bigg)\nonumber \\
 & - & \frac{1}{\rho r^{2}\sin\theta}\frac{\partial}{\partial\mu}\sin^{2}\theta\bigg(T_{\theta\phi}\left(\Omega_{\odot}+\delta\Omega,\Omega^{*},\beta\right)\nonumber \\
 & - & T_{\theta\phi}\left(\Omega_{\odot},\Omega^{*},\beta=0\right)-\frac{B}{4\pi r\sin\theta}\frac{\partial A}{\partial r}\bigg),\nonumber 
\end{eqnarray}
where $\beta={\displaystyle
  \frac{\left|\overline{B}\right|}{\sqrt{4\pi\overline{\rho}u'^{2}}}}$
characterizes the strength of the magnetic field, $\Omega_{\odot}=\Omega_{0}^{\odot}f\left(r,\theta\right)$,
$f\left(r,\theta\right)$ the angular velocity profile given by
helioseismology, $T_{\left(\theta,r\right)\phi}\left(\Omega_{\odot},\Omega^{*},\beta=0\right)$
are the turbulent stresses (see, Eqs(\ref{eq:trf},\ref{eq:ttf})) in initial
state when there is no large-scale magnetic field. 
 The same is applied for the star rotating with the different
period than the Sun. We assume that the sum of the radial turbulent
and the Maxwell stresses is zero at the boundaries. This guarantees
the conservation of the angular momentum in the model. 
{In the absence of the large-scale magnetic field any perturbation
$\delta\Omega$ evolves to zero.}  The details of
the turbulent stresses expressions are given in Appendix.
\begin{table*}
\begin{minipage}{180mm}%
 \protect\caption{Parameters of the models and results. The <<$\pm$>> means inclusion or exclusion
of the non-linear effect from the model. The last three columns show the results
for the set of rotation rates $\left\{ 0.85,1,1.25,1.5,1.6,1.75\right\} \times\Omega_{0}$,
where $\Omega_{0}=2.8\times10^{-6}$ s$^{-1}$. {They are, the total magnetic flux
which includes the unsigned magnetic flux of the toroidal field
integrated over the whole convection zone,} the period
of the magnetic cycle, the total Poynting flux from subsurface shear layer. The multiply periods
( in case $1.75\times\Omega_{0}$ ) are shown are shown in brackets.}
\begin{tabular}{@{}llrrlll@{}}
\hline 
Model  & $\overline{\chi}$  & ${\displaystyle
  \frac{\partial\delta\Omega}{\partial t}}$
  & Buoyancy & Flux {[}$10^{24}$Mx{]}  & $P_{cyc}${[}Yr{]}  & Poynting Flux,~{[}$10^{-3}F_{\odot}${]}\\
\hline 
D0  & -  & -  & +  & 0.9,5.2,7.1, 8.1,8.4,9.1  & 12.,8.,6., 4.8,4.4,(3.2,4.7)  & 0.9,1.5,2.3, 3.0,3.3,3.9\\
D1  & +  & -  & +  & 0.7,2.2,4.3, 5.6,6.1,7.3  & 12.,10.5,8.0, 5.6,5.0,(4.6,4.9)  & 0.03,0.3,0.8, 1.1,1.3,1.8\\
D2  & -  & +  & +  & 0.8,4.9,6.9, 8.1,8.4,9.1  & 12.,8.3,6.5, 5.4,4.9,4.4  & 0.03,1.3,2.2, 3.0,3.3,3.9\\
D3  & +  & +  & +  & 0.7,1.9,4.0, 5.4,6.0,6.5  & 12.,10.4, 8.1, 6.3,6.0,(5.4,5.7)  & 0.03,0.3,0.7, 1.1,1.3,1.4\\
B0  & -  & -  & -  & 4.4,6.5,10.4, 15., 17., 20.  & 12.1,13.,14., ~~~14.8, 15.4,16.5  & \\
B1  & -  & -  & +  & 2.6,3.0, 4.1, 5.0, 5.34,5.8  & 11.5,11.8,12.4, 12.7,13.0,13.4  & \\
B2  & +  & -  & +  & 1.1,1.4,1.7 2.1,2.2,2.3  & 10.7, 10.9,11.1, 11.3,11.6,12.  & \\
B3  & -  & +  & +  & 2.2,2.9,4.0, 4.9,5.2,5.6  & 11.5,12.1,12.6, 13.4, 13.7,14.0  & \\
B4  & +  & +  & +  & 1.1,1.4, 1.7, 2.1,2.2,2.3  & 10.6,10.7,11., 11.3,11.5,11.9  & \\
\hline 
\end{tabular}
\end{minipage}
\end{table*}

\begin{figure*}
\includegraphics[width=.99\textwidth]{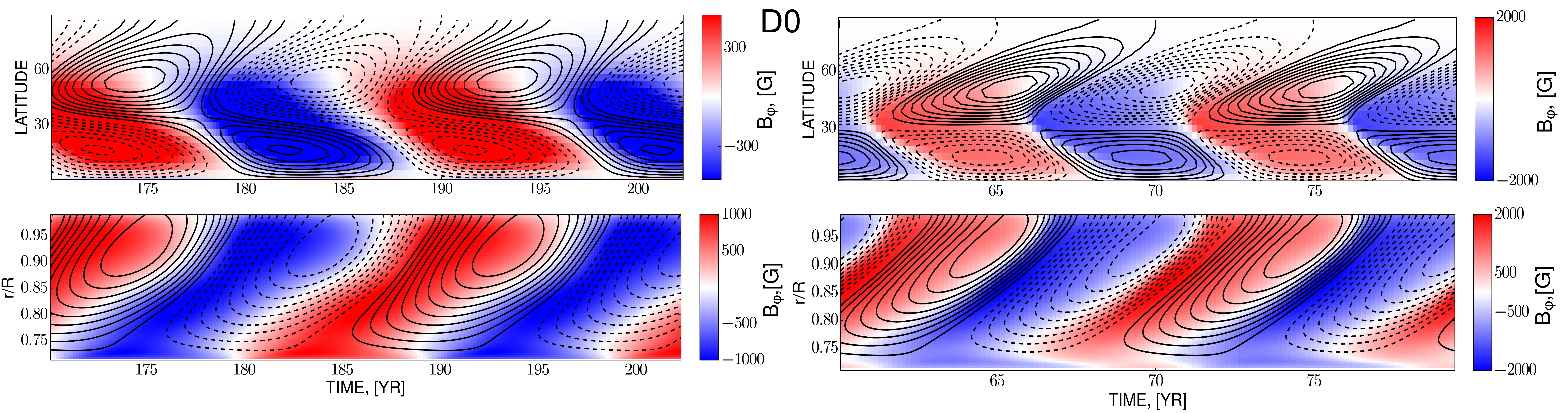}
\protect\caption{The time-latitude diagrams (top) for the radial magnetic field (contours)
at the surface and the near-surface toroidal magnetic(image) field
for the model D0 for the periods of rotation 25 (left) and 15 (right)
days. The bottom line shows the same for the time-radius variation
at $30^{\circ}$ latitude. }
\end{figure*}

\begin{figure*}
\includegraphics[width=0.99\textwidth]{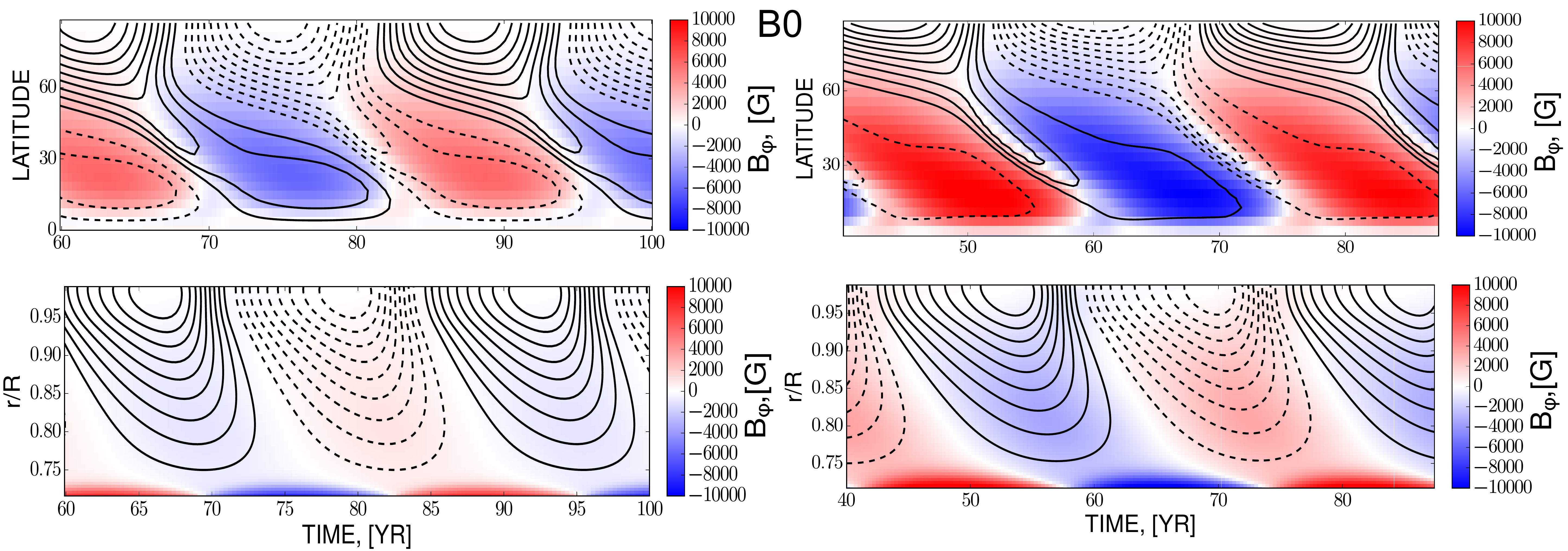}
\protect\protect\caption{The same as Figure 1 for the model B0. The toroidal field in the
time-latitude diagram is taken from the bottom of the convection zone. }
\end{figure*}

\section{Results}

The study employs the same set of the basic parameters as in our
previous paper (see, e.g. \citealp{pip13M,2014ApJ_pipk}).  Summary of
the dynamo models is presented in the Table 1. {The model  parameters are chosen
to reproduce the basic properties 
 of the solar cycle, such as the cycle period of $\sim
11$ Yr, the magnitude of the polar field, $\sim 1$G, the
total magnetic flux in the solar convection zone produced by the
dynamo, $\sim10^{24}$Mx, the
time-latitude diagrams for the near-surface toroidal and radial
magnetic fields (including the phase relation between them) as discussed in our previous
papers (see, e.g, \citealt{2014ApJ_pipk}).} 
For the B-type model we use parameters to make results 
for the solar case  close to that discussed by \citet{2011AstL...37..286K}.

The numerical integration is carried out in latitude from pole to
pole, and in radius from $r_{b}=0.715R_{\odot}$ to
$r_{e}=0.99R_{\odot}$. The numerical scheme employs the pseudo-spectral approach
for the numerical integration in latitude and the finite second-order
differences in radius. The initial field was a weak dipole type poloidal
magnetic field. The  numerical scheme preserves the parity of
the initial field unless there is a real parity braking process,
e.g., associated with non-symmetric, relative to the equator, perturbations
of the dynamo parameters. The initial conditions represent a weak dipole field with strength 0.01G.

For each rotation rate from the set $\left\{ 0.85,1,1.25,1.5,1.6,1.75\right\} \times\Omega_{0}$,
where $\Omega_{0}=2.87\times10^{-6}$s$^{-1}$ we study the different
nonlinear dynamo regimes. The models D0 and B0 is related to the
dynamo with the ``algebraic'' quenching of the $\alpha$-effect. The
model D0 includes the buoyancy effect as well.
In increasing the number of the model we include the other nonlinear
effects, such as, the magnetic helicity and
the magnetic feedback on the differential rotation, see details in
the Table 2. The realistic model should include all those effects into
account. 
{The output of the models which we will use for comparison
with observations includes, the total magnetic flux, which is integrated
unsigned magnetic flux of the toroidal field over the bulk of the convection zone, the period
of the magnetic cycle, the total Poynting flux from subsurface shear
layer.} 

\subsection{Spatial structure of dynamo waves}

Figure 1 shows the time-latitude diagrams for the radial magnetic
field at the surface and the near-surface toroidal magnetic field
for the model D0, which takes into account the nonlinear
$\alpha$- effect, and the magnetic buoyancy for the rotation periods of 25 and 15 days.
Figure 2 shows the same as Figure 1 for the model B0. The butterfly
diagrams of the model B0 for the period of rotation 25 days is rather
similar to those by \citet{2011AstL...37..286K}. {They noticed
that in the model with the non-local $\alpha$-effect  the dynamo wave does not follow
to the Parker-Yoshimura rule. Thus, in their model the equatorward propagation of the
dynamo wave is likely due to the joint action of the latitudinal shear and the strong
diamagnetic pumping effect to the bottom of the convection zone.}

The Figures 1,2 illustrate the main difference between the models
with the ``local'' and non-local $\alpha$- effect. 
The radial propagation of the dynamo wave is outward
for the D-type models and it is opposite for the B-type models. The downward propagation of the dynamo wave is a typical
feature of the Babcock-Leighton types dynamo models (see, e.g., \citealt{dc99}).
{ The downward propagation of the dynamo waves drives the magnetic field
from the high latitudes at the top to  the low latitudes at the bottom
of the convection zone where the turbulent magnetic  diffusivity is small. The amplitude of the turbulent diffusion
is the key parameter that determines the period of the dynamo
cycle. The increase of the rotation rate amplifies the  concentration
of the dynamo wave  to the bottom of the convection zone. This
likely } results to increase of the dynamo
period. {Also, in the given examples of the
B-type models, the maxims of the polar magnetic field corresponds to
the maxims of the toroidal field at the low latitudes (cf, with the
D-type models).} 
\begin{figure*}
\includegraphics[width=0.9\textwidth]{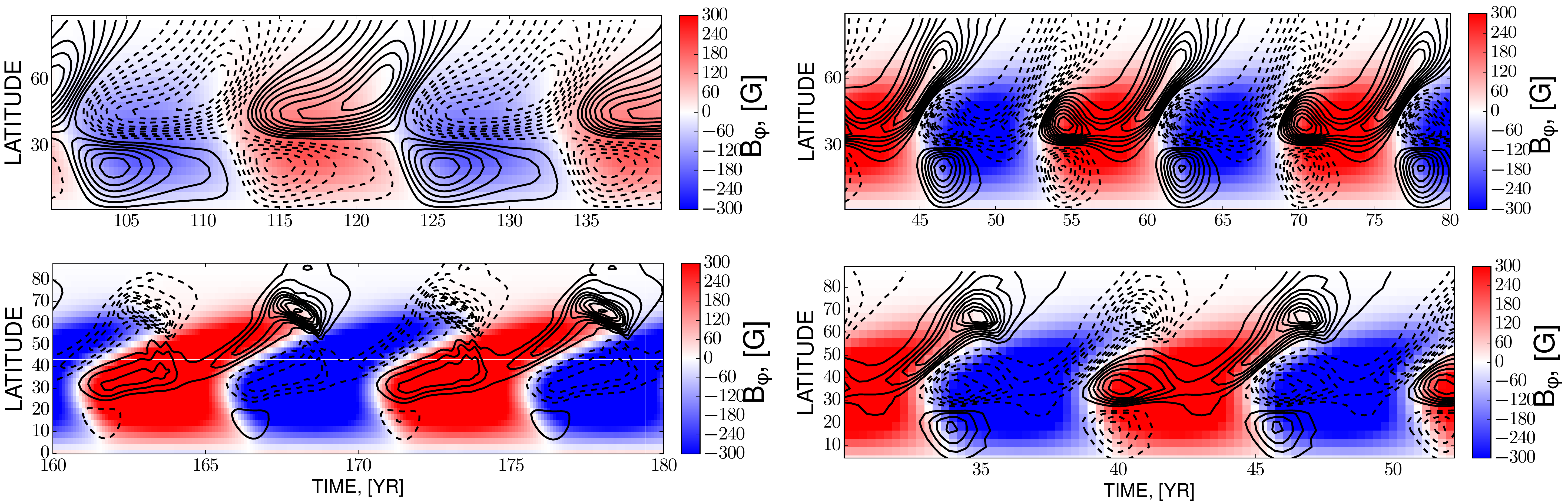}
\protect\protect\caption{The time-latitude diagrams for the radial magnetic field (contours)
at the surface and the near-surface toroidal magnetic(image) field
for the model D1 for the set of the periods of rotation 25.3, 20.3,16.9
and 14.5 days clockwise starting from the up left.}
\end{figure*}

The nonlinear effects due to magnetic buoyancy, magnetic helicity
conservation results to saturation both
the D-type and B-type dynamo models. 
Recently, \citet{2014arXiv1412.0997B} re-considered this
effect for the two different kinds of the dynamical quenching in the
B-type models. Here
we confirm their results for the axisymmetric spherical dynamo models.
 Results in the Table 1 (see the
B1) shows that the magnetic buoyancy quenches the magnetic flux produced
by the dynamo by factor 2 (for the rotation period 25 days) to factor
4 (for the rotation period of 15 days). The magnetic buoyancy is less
essential for the D-type models \citep{1993A&A...274..647K}. 

Quenching of the dynamo by the magnetic helicity conservation make
a little change to the structure of the butterfly diagram of the
B-type model. The changes in the D-type models are considerable.
They are illustrated in the Figure 3. We find that the polar branch
of the toroidal magnetic field becomes more and more pronounced when
the period of the rotation is decreased. Also the character of the
polar reversals is eventually changed from the smooth and monotonic
kind as in the case of the 25-days period to the serge like pattern
when the period of rotation is 14.5 days.

\begin{figure*}
\includegraphics[width=0.95\textwidth]{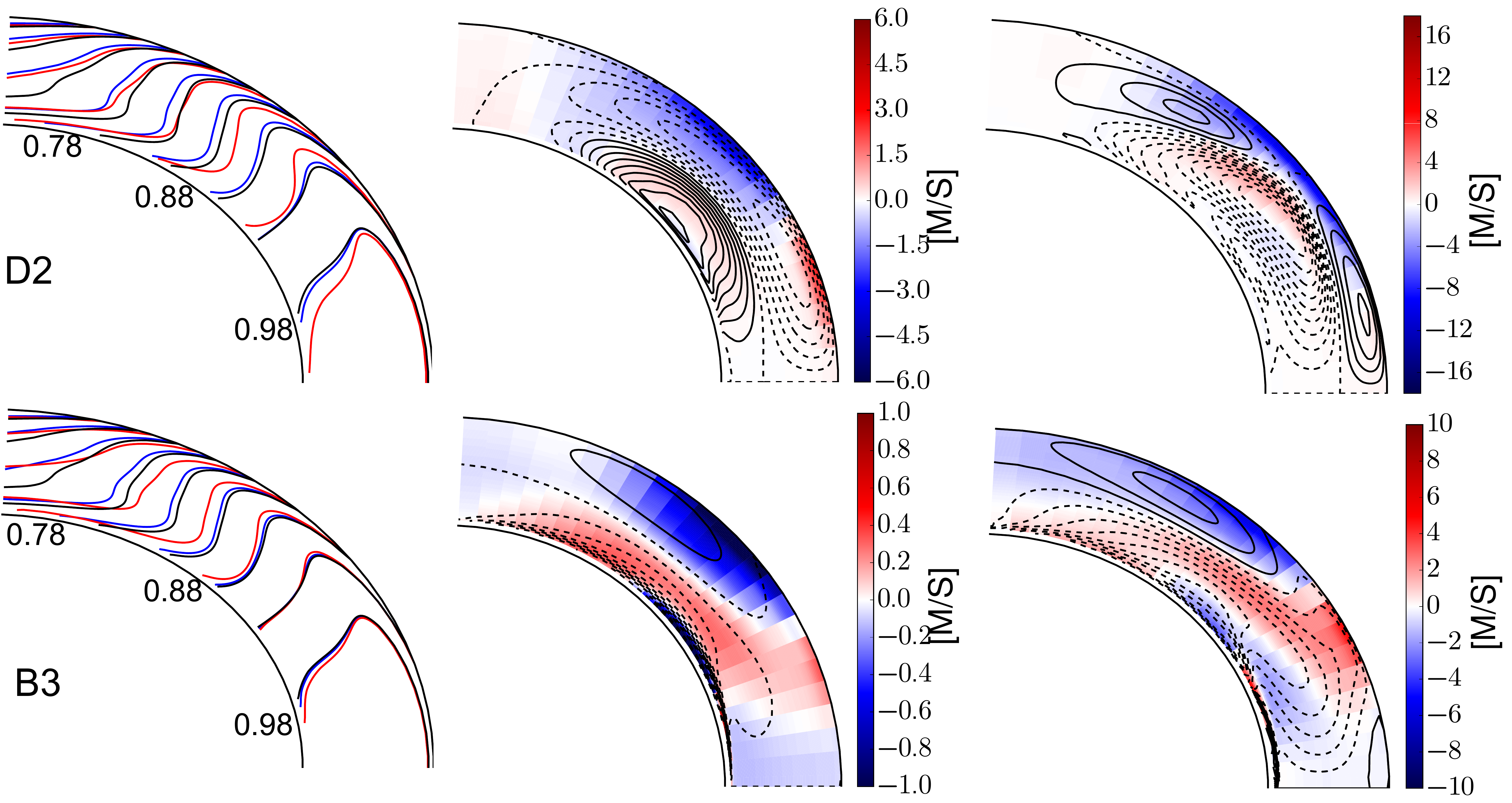}
\protect\protect\caption{Top line: left panel shows the isolines of the mean distribution of
the angular velocity in the star in the range
$[0.63-0.98]\times$$\Omega\left(r,\theta\right)/\Omega_{\boldsymbol{\star}}$
($\Omega_{\boldsymbol{\star}}$ - rotation rate of  a star)
with particular levels at 0.78,0.88 and 0.98
for the periods of rotation 25.3(blue color) and 14.5 days (red color),
black color shows the unperturbed distribution; middle panel shows
snapshots of the magnetic field distribution (contours) and the torsional
oscillation (color image) for the model D2 with the period of rotation
25.3 days; the right panel shows the same as the middle panel for
the period of rotation 14.5 days. The bottom line shows the same as
the top line for the model B3.}
\end{figure*}

\subsection{Magnetic feedback on the differential rotation}
\begin{figure}
\includegraphics[width=0.8\columnwidth]{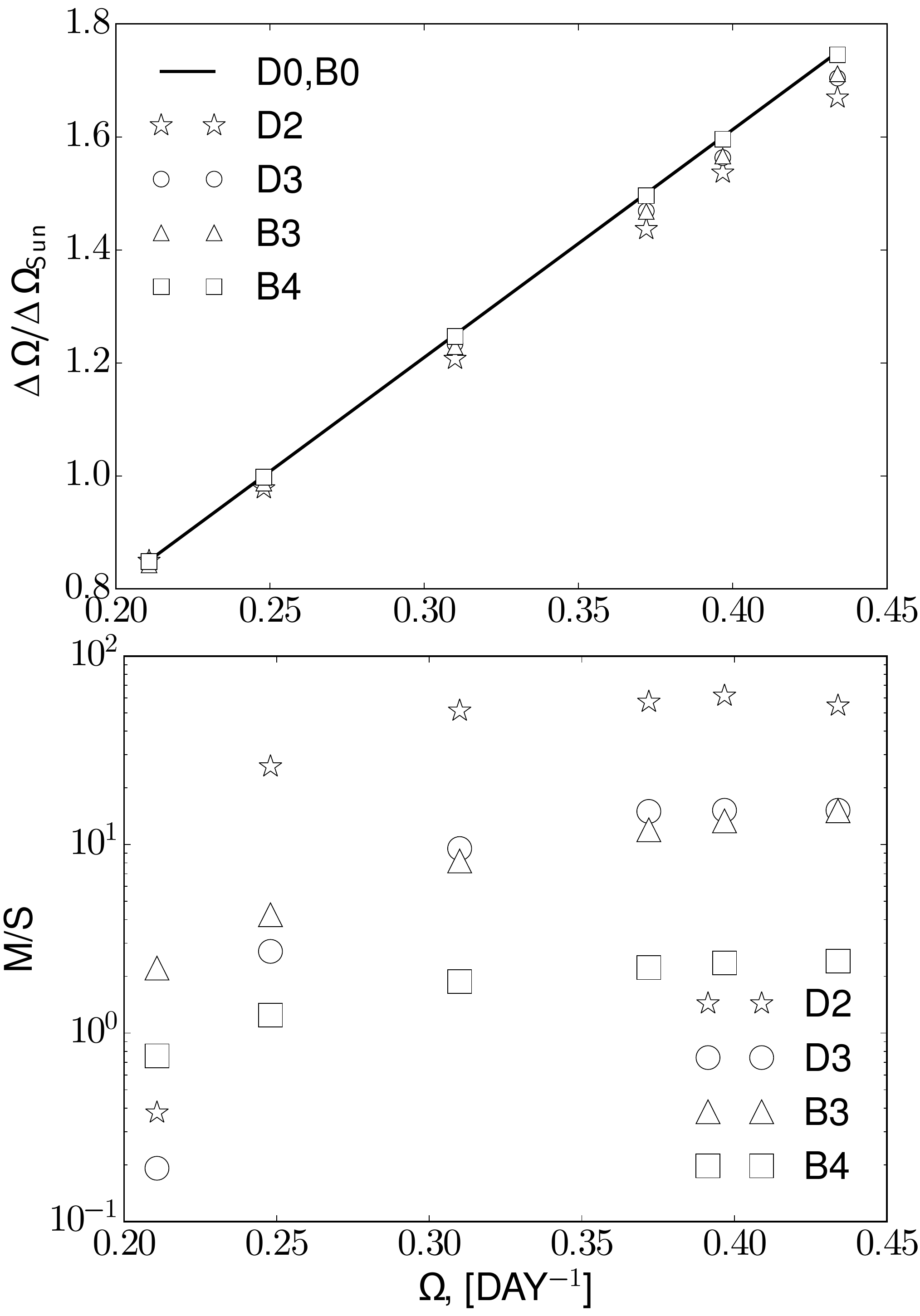}
\protect\protect\caption{Top panel shows the mean
  surface differential rotation at the stationary state of the
  magnetic activity evolution, the black line shows the kinematic
  case. The bottom panel shows { the maximum amplitude of the zonal velocity perturbation at latitude of
30$^{\circ}$ north on the surface for the models D2, D3, B3 and B4.} }
\end{figure}
\begin{figure}
\includegraphics[width=0.8\columnwidth]{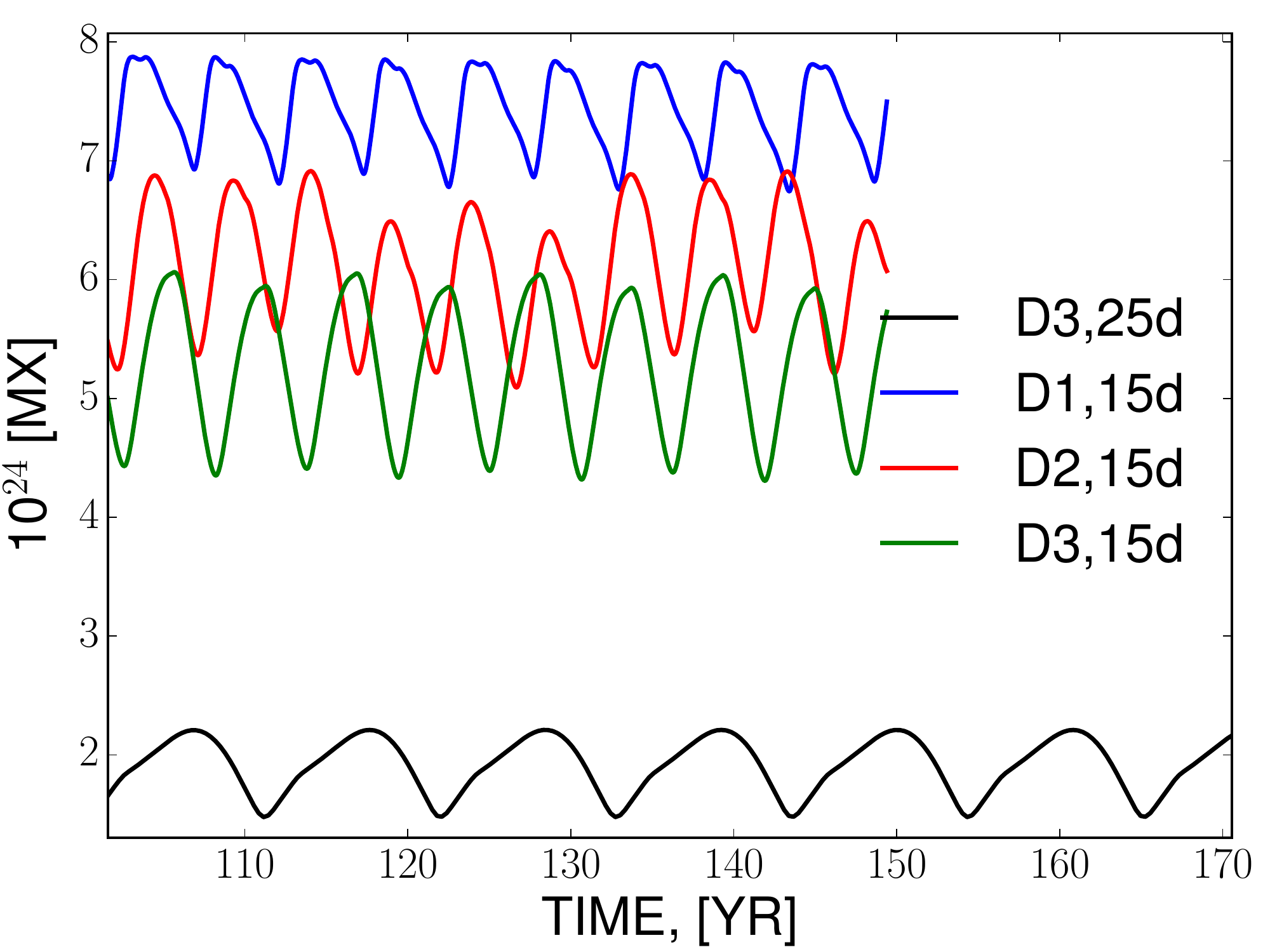}
\protect\protect\caption{Variation of the total flux in D-type models}
\end{figure}

The further quenching of the dynamo process can be seen when we take
into account the back reaction of the large-scale magnetic field on
the angular momentum transport. It is found that  the D- and B-type
dynamo models produce the different effect on the angular momentum transport inside the convection
zone. In the B-type model
the torsional oscillations are concentrated to the bottom of the dynamo
domain. Note that the situation can be different for the B-type
model with regards for the meridional circulation and the magnetic
backreaction on the heat transport (see, \citealt{2006ApJ...647..662R}).

Figure 4 shows the snapshots of the torsional oscillations and the
mean states of the angular velocity distributions for the 
periods of rotation of 25.3 and 14.5 days for the D2 and B3 models. { The
  torsional oscillations are determined 
as the difference between the instantaneous and the mean profiles
 of the angular velocity for the stationary stage of the magnetic
activity evolution.} The D2 and
B3 models show the stronger deviation of the
rotation profiles from the unperturbed state than the models D3 and B4. The magnetic helicity
conservation quenches the amplitude of the torsional oscillations by
means of damping the magnitude of the dynamo generated magnetic fields. For
the D2 and B3 type models the distribution of the angular velocity
deviates strongly from the radial-like profiles for the star rotating
with the period of 14.5 days (see, e.g., Figure 4 (top,left). In the
model D2 and B3 the rotation profiles in the mid part of the convection
zone becomes more and more cylinder-like when the period of the rotation
decreases.
 Figure 5 shows results for the surface differential rotation and the magnitude
of the torsional oscillations in the dynamo modes D2, D3, B3 and
B4. {The magnitude of the torsional oscillations is determined by the
maximum amplitude of the zonal velocity perturbation at latitude of 
30$^{\circ}$ north on the surface in the stationary
state of evolution.}

We find that in the  model D2 the magnetic activity reduces
the latitudinal shear at the surface by 10\% of the
$\Delta\Omega_{\odot}$,
 and  the reduction is about 5\% for the fully nonlinear dynamo model
 D3. In the B-type models the reduction
 of the latitudinal shear is about 5 \% of
the $\Delta\Omega_{\odot}$ in the case B3. In the fully
nonlinear B-type dynamo model the change  of the latitudinal shear is relatively small.
It is less than 1 \% of the $\Delta\Omega_{\odot}$. The magnitude
of the torsional oscillations behave similarly. The model D2 with the
period of rotation 14.5 days shows the torsional oscillations of
magnitude about 70 m/s at the surface.

\subsection{Dynamo amplitude and period vs period of rotation}

We find that for the fast rotation case the nonlinear dynamos of D-type
can show the multiply periods. However, in the given range of the
rotational periods and for the given choice of the dynamo parameters, the regime of the multiply dynamo periods
is not well developed. Figure 6 illustrates variations
of the total unsigned magnetic flux of the toroidal field generated by the dynamo in the model D3
for the rotational periods 25.3 and 14.5 days and for the models D1
and D2 for the rotational period 14.5 days. For the period of rotation
14.5 days the D1 model for the shows the <<long-term>> variations
with period about 5-6 of the basic periods. { The long-term variations
are damped by the nonlinear $\Lambda$- quenching effects and the large-scale
Lorentz forces. It seems that the model D3 (with the rotational period 14.5 days)
is near the threshold of the nonlinear regime with the <<long-term>>
variations.} This question should be studied further.

\begin{figure*}
\includegraphics[width=.8\textwidth]{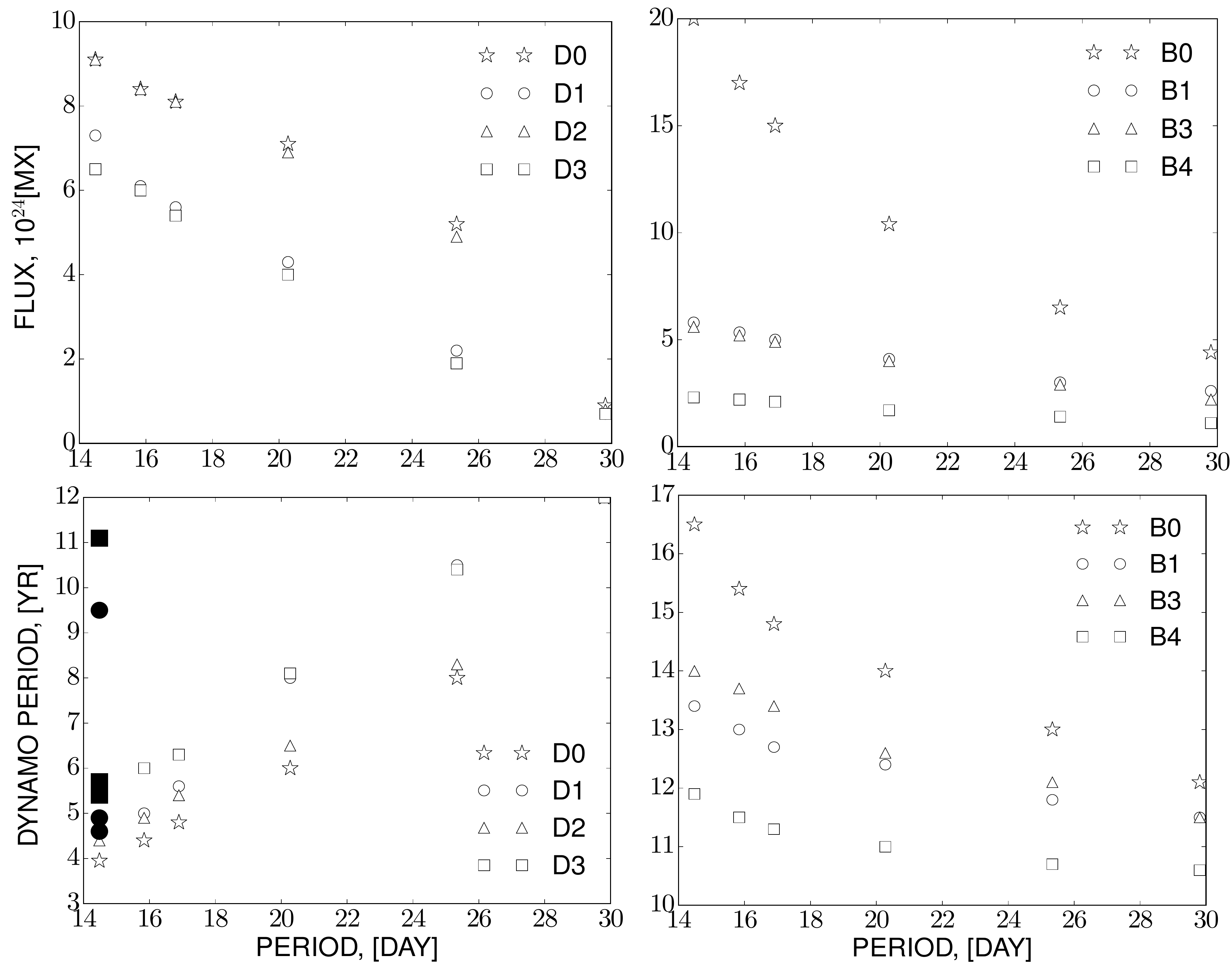}
\protect\protect\caption{The top line show the dependence of the total magnetic flux generated
by the dynamo versus the period of rotation. The bottom line shows
the variations of the dynamo period. The filled symbols shows results
for the models D1 and D3 with the multiply periods. In those cases,
the results for the sum of the principal periods are shown as well. }
\end{figure*}

Figure 7 illustrates our findings about dependence of the magnetic
cycles parameters, such as the magnitude of the total magnetic flux
and the dynamo period, on the period of rotation of the star.
{ The models D1 and D3 have the weaker dynamo than the models D0 and D2.
The model with the non-local $\alpha$ effect, the B0, has the strongest toroidal field for the set of our
dynamo models. Surprisingly, that the amplitude of the toroidal magnetic flux in the
models B1 and B3  is rather similar to the models D1 and D3. This
tells us that the magnetic buoyancy quenches the B-types dynamo models in a
more efficient way than for the D-type. Also, the total magnetic flux in the fully nonlinear B-type
model is about factor 3 smaller than in the D-type.}

The D-type models shows the decrease of the dynamo period with the
decrease of the rotational period of the star. The period of magnetic
cycles changes from about 11 years for the solar case to the value
which is about 3 year for the star rotating with the period about
15 days. The opposite tendency is found for the B-types dynamo models,
where the dynamo period increases from 11 to 17 years when the
rotational period decreases.

{The Table 2 gives results for the mean Poynting magnetic
  energy flux from the subsurface shear
 layer, $S_M$, for the D-type models. In the model we use the special
 boundary conditions, the Eq.(5)  which allow the penetration of the
 toroidal field to the surface. In the linear case, this condition
 gives $S_M\sim B_s^2$, where $B_{s}$ is the  toroidal
 magnetic field strength in the near-surface layer.
 In simulations it is found that the increase of the $S_M$ is
 gradually slow down when the  rotation rates increase.
Thus we can conclude that in the nonlinear dynamo the
generated Poynting flux increases as $S_M\sim B_{s}^{2-\delta}$ where the
$0 <\delta<1$.}
This was  predicted by \citet{1995A&A...297..159K} (cf., to
\citealt{2015MNRAS.446L..51B}) for the dynamo models with magnetic
helicity conservation as the principal dynamo non-linearity. It is
confirmed by \citet{vid14MNR} who found $\delta\approx1.61$ for the
X-ray flux for magnetic activity in solar type stars. 
The flux reaches the magnitude $10^{-3}F_{\odot}$, where
$F_{\odot}={\displaystyle \frac{L_{\odot}}{4\pi R_{\odot}^{2}}}$. 
The result is in the agreement with the observational findings \citep{2006ARep_kazova}.
We postpone the detailed study of the Poynting flux in the nonlinear
dynamo for the future.

{In our calculations we compute the averaged surface
magnetic field, as well. The results for this proxy of the stellar magnetic
activity were published recently by \cite{vid14MNR}. For the given
range of the rotation rates both the D- and B-types of the dynamo
models reproduce the observational finding reasonably well. We
postpone a more detailed comparison of these results with observations
for a future.}

{It is noteworthy that the feedback of the magnetic helicity
quenching on the large-scale dynamo depends on amount of
the magnetic helicity in the dynamo region. This is partly controlled by the effective magnetic
Reynolds number, $R_m$. The decrease of the $R_m$ simulates
the increase of the helicity loss. It is found that for the
$R_m=10^2$ the model D3 produces nearly the
same amount of the total magnetic flux as the model D2. 
The similar effect can be expected for the B-type
models. The increase of the turbulent diffusivity of the magnetic
helicity will produce the similar effects as the decrease of the
$R_m$ (see \citealp{hub-br12}).
It is likely that the helicity loss is modulated by the
magnetic activity, e.g., by modulation of the coronal activity. Thus,
the realistic dynamo model have to capture the basic mechanisms of the
magnetic helicity loss, e.g., by means of including the stellar
corona  in simulations (see, \citealp{bl-br2003,warn2011}).}
  
{ The D-type model was calibrated against the solar case.
The eigenvalue analysis for the D-type models (see, Pipin 2012, and
references therein) shows that variations of the free parameters, like the $C_{\delta}$ (the
$\Omega\times J$ effect), the $C_v$ (stratification of the turbulent
diffusivity) can produce a dynamo with steady and non-oscillating large-scale
magnetic field. The changes from the oscillating to steady dynamo
regimes with the increase of the rotation rates could happen in the
D-type dynamo models for the range  of the
parameters $C_{\delta,v}$ different than we use in the paper. The
nonlinear dynamo models in these cases should be studied separately.}

\section{Discussion and conclusions}

In the paper we have studied the nonlinear effects in the large-scale
magnetic solar-type dynamo on the parameters of the dynamo cycles
for the range of the rotation periods from 14.5 to 30 days. This a
preliminary study. The dynamo model lacks the self-consistent description
of the angular momentum in the convection zone of a star. We also
exclude the evolutionary changes of the stellar convection zone structure,
which could occur together with the angular momentum loss. The model
also lacks the effects of the meridional circulation to the dynamo
action, which is found in the direct numerical simulations \citep{miesch11,guer2013,2014A&A...570A..43K}
and it is extensively used in the advection B-type solar dynamo models
\citep{raey}. Therefore I consider the results of the paper as preliminary.

We find that the D-type dynamo models (the dynamo distributed over
the convection zone) satisfactory explains the dependence of the dynamo
magnitude and the dynamo periods on the periods of rotation of the
star (see, e.g., \citealt{soon94,2011IAUS_saar}). {These properties
are expected from the qualitative predictions made for linear 
and weakly-nonlinear Parker-type dynamo
models \citep{ku98}. For the Sun, the relation between the magnitude
of the solar cycle and its period is a part of the Waldmeier's
relations \citep{w36}.}
 The B-type models
fails to explain the latter fact. In mean-field models of the stellar
dynamo it was seen earlier by \citet{2010A&A...509A..32J} and \citet{2014ApJ...791...59K}.
Here we confirm this effect on the B-type model without meridional
circulation. Note, that the increase of the dynamo period
with increase of magnitude of dynamo generated magnetic field was reported earlier by
\citet{1995A&A...296..557R} for the dynamo model in the overshoot layer
below the convection zone.
In their case the dynamo wave propagate inward to the bottom of the
overshoot layer, see Fig.4 (right) in \citep{1995A&A...296..557R}.
The similar is found in our study for the B-type  model. 
This proves the generic character of this effect for this
type of the large-scale dynamo. Figure 2 shows that in the B-type
models the dynamo wave propagates to the bottom of the convection
zone. Which is promoted by the diamagnetic pumping. The effect can
be amplified in the model with the meridional circulation. The concentration
of the dynamo wave to the bottom of the convection zone, where the
magnetic diffusivity is low, is amplified for the fast rotating star.
This enlarges the amplitude of the dynamo and the dynamo period as
well. The certain set of the multiply meridional circulation cells
can help to solve this issue \citep{2010A&A...509A..32J}. However, it
is not clear if such a set of the meridional circulation cells is
consistent with the angular momentum transport.

{It is found that the magnetic helicity conservation is the most efficient
mechanism  for the nonlinear quenching of the large-scale dynamo.  
 We confirm the results about magnetic helicity quenching of the B-type dynamos which
were reported earlier by \citet{2007NJPh....9..305B} and \citet{2014arXiv1412.0997B}.
Results for the given range of the rotational periods does not
give enough confidence  to conclude if there is a saturation level is
reached  in the B-type dynamos at the rotation period 15 days. An increase of the
generated magnetic flux with the increase of the rotation rate remains
constant for the all D-type dynamos.
It is noteworthy, that the model of the magnetic helicity loss should be elaborated
further for a better confidence in comparing the results of the
mean-field dynamo models with observations.}

The large-scale Lorentz force and the $\Lambda$- quenching reduce
the surface differential rotation and produce the torsional oscillations.
The effects are stronger in the D-type dynamo models than in the B-type
models. This seems due to the distributed character of the magnetic
feedback
 on the angular momentum balance for the dynamo operating in
the whole convection zone. It is found that the magnetic
backreaction on the surface differential rotation can be about 10 \% of
the solar value. Thus in the given interval of the rotational period
the magnetic saturation of the differential rotation by the dynamo
(see, e.g., \citealt{2011IAUS_saar}) is not significant. Following
to hints from observations it could be found for the higher rotation
rates.

{Analysis of observations of stellar magnetic activity (see, e.g.,
\citealt{soon93,Baliunas1995,1999ApJ_sa_br,2007ApJ_bohm}) reports
the multiply branch of the dynamo activity for the cool stars. Results
show  the multiply periods on the  Sun's analog rotating with
period about 15 days. The regime with multiply periods was reported in
the solar-type dynamo models when the nonlinear system passes from the
regular to chaotic behaviour (see, e.g.,
\citealp{bran89,cov98,tob1996,p99}).
 In our set of the D-type models, the most realistic are the D2 and D3. Variation of the
rotation rate  in those models gives the different distribution of the
periods. In reality,  the magnetic helicity loss (see the end
of subsection 3) we would give the mix of the distributions which are
produced by the models  D2 and D3.   Therefore the given results do not show
the  distinct multiply branch of the dynamo activity. This interesting
question will be addressed in the future studies.}

{I summarize the main results as follows. The nonlinear distributed
mean-field  solar dynamo model is calibrated versus the solar case. It has the basic properties
of the solar cycle such as the cycle period, $\sim 11$Yr, the total
magnetic flux produced in the dynamo, $\sim 10^{24}$Mx and the time-latitude diagrams
for the large-scale magnetic field, in a good agreement with
observation. The same model satisfactory reproduces the main properties of the
stellar magnetic cycles on the Sun's analogs for the range of
rotational periods of 15-30 days.}   

\subsection*{Acknowledgements} {The work is supported by RFBR under grants
  14-02-90424, 15-02-01407 and the project  II.16.3.1 of ISTP SB RAS. I'm grateful to anonymous referee for
  comments and suggestions. I also thank D. Moss, A.G. Kosovichev and K.M. Kuzanyan  
for a critical reading of the manuscript and helpful comments.}


\section{Appendix A}

\setcounter{equation}{0}

\global\long\def\theequation{A\arabic{equation}}

\subsection{The $\boldsymbol{\mathcal{E}}$}

This section of Appendix describes the parts of the mean-electromotive
force which contribute in the Eq.(\ref{eq:EMF-1}). 
\subsubsection{The anisotropic diffusion}
The anisotropic diffusion tensor $\eta_{ijk}$ was derived in \citep{pi08Gafd} (hereafter P08) and \citep{2014ApJ_pipk}.
 The $\eta_{ijk}$ takes into account the generation of the magnetic fields by the $\Omega\times J$ effect
\citep{rad69}, as well. The $\eta_{ijk}$ reads  
\begin{eqnarray}
\eta_{ijk} & = & 3\eta_{T}\bigg\{\left(2f_{1}^{(a)}-f_{2}^{(d)}\right)\varepsilon_{ijk}+2f_{1}^{(a)}e_{i}e_{n}\varepsilon_{jnk}\label{eq:diff}\\
 & + & \frac{a}{3}\phi_{1}\left(g_{n}g_{j}\varepsilon_{ink}
-\varepsilon_{ijk}\right)\bigg\} \nonumber\\
&+&3\eta_{T}C_{\delta}f_{4}^{(d)}e_{j}\left\{
  \tilde{\varphi}_{7}^{(w)}\delta_{ik}
+\tilde{\varphi}_{2}^{(w)}\frac{\overline{B}_{i}\overline{B}_{k}}{\overline{B}^{2}}\right\} ,\nonumber 
\end{eqnarray}
where $\mathbf{e}={\displaystyle \boldsymbol{\Omega}/\Omega}$ is
the unit vector along the rotation axis and $\mathbf{g}$ is the unit
vector in the radial direction, $a$ is the parameter of the turbulence
anisotropy, $\eta_{T}$ is the magnetic diffusion coefficient. The
components of the $\eta_{ijk}$ depend on the Coriolis number $\Omega^{*}=4\pi{\displaystyle \frac{\tau_{c}}{P_{rot}}}$,
where $P_{rot}$ is the rotational period, $\tau_{c}$ is the convective
turnover time. The quenching functions $f_{1,2}^{(a,d)}$ and $\phi_{1}$
are
\begin{eqnarray*}
f_{1}^{(a)} & = & \frac{1}{4\Omega^{*\,2}}\left(\left(\Omega^{*\,2}+3\right)\frac{\arctan\Omega^{*}}{\Omega^{*}}-3\right),\\
f_{2}^{(d)} & = & \frac{1}{\Omega^{*\,2}}\left(\frac{\arctan\Omega^{*}}{\Omega^{*}}-1\right),\\
\phi_{1}\!\! & = &\!\! -\frac{1}{24\Omega^{\star2}}\left(2\log\left(1+4\Omega^{\star2}\right)+4\log\left(1+\Omega^{\star2}\right)\right.\\
 & + &
 \left.\left(1-4\Omega^{\star2}\right)\frac{\arctan\left(2\Omega^{\star}\right)}{\Omega^{\star}}\right.\\
&+&\left. 4\left(1-\Omega^{\star2}\right)\frac{\arctan\left(\Omega^{\star}\right)}{\Omega^{\star}}-6\right).
\end{eqnarray*}
The last term in the Eq.(\ref{eq:diff}) is a contribution of the
$\Omega\times J$ effect with a free parameter $C_{\delta}$ and the
quenching functions of the magnetic field and the Coriolis number,$\tilde{\varphi}_{2,7}^{(w)}\left(\beta\right)$
and $f_{4}^{(d)}\left(\Omega^{*}\right)$: 
\begin{eqnarray*}
\tilde{\varphi}_{2}^{(w)}\!\! &\! =\! &\!\! -\frac{5}{192\beta^{5}}\left(\!3\left(16\beta^{4}\!-\!5\right)\frac{\arctan\left(2\beta\right)}{2\beta}-5\left(4\beta^{2}-3\right)\right),\\
\tilde{\varphi}_{7}^{(w)} & = & \frac{5}{192\beta^{4}}\left(3\left(16\beta^{4}-1\right)\frac{\arctan\left(2\beta\right)}{2\beta}-\left(4\beta^{2}-3\right)\right),\\
f_{4}^{(d)} & = & \frac{1}{6\Omega^{*\,3}}\left(\left(2\Omega^{*\,2}+3\right)-3\left(\Omega^{*\,2}+1\right)\frac{\arctan\left(\Omega^{*}\right)}{\Omega^{*}}\right).
\end{eqnarray*}
Note in the notation of P08 $\tilde{\varphi}_{2,7}^{(w)}\left(\beta\right)=\frac{5}{2}\varphi_{2,7}^{(w)}\left(\beta\right)$
and $\beta={\displaystyle \frac{\left|\overline{B}\right|}{\sqrt{4\pi\overline{\rho}u'^{2}}}}$. 

\subsubsection{The $\alpha$-effect}
The $\alpha$ effect takes into account the kinetic and magnetic helicities,
\begin{eqnarray}
\alpha_{ij} & = & C_{\alpha}\sin^{2}\theta\psi_{\alpha}(\beta)\alpha_{ij}^{(H)}\eta_{T}+\alpha_{ij}^{(M)}\frac{\overline{\chi}\tau_{c}}{4\pi\overline{\rho}\ell^{2}}\label{alp2d-1}
\end{eqnarray}
where $C_{\alpha}$ is a free parameter, the $\alpha_{ij}^{(H)}$
and $\alpha_{ij}^{(M)}$ represent the kinetic and magnetic helicity
parts of the $\alpha$-effects, respectively, $\overline{\chi}$-
is the small-scale magnetic helicity, the $\ell$ is the typical length
scale of the turbulence, and $\bar{\rho}$ is the mean density. The $\alpha_{ij}^{(H)}$ reads, 
\begin{eqnarray}
\alpha_{ij}^{(H)} & = & \delta_{ij}\left\{
  3\left(f_{10}^{(a)}\left(\mathbf{e}\cdot\boldsymbol{\Lambda}^{(\rho)}\right)
+f_{11}^{(a)}\left(\mathbf{e}\cdot\boldsymbol{\Lambda}^{(\eta)}\right)\right)\right\} \label{eq:alpha}\\
 & + & e_{i}e_{j}\left\{ 3\left(f_{5}^{(a)}
\left(\mathbf{e}\cdot\boldsymbol{\Lambda}^{(\rho)}\right)+f_{4}^{(a)}\left(\mathbf{e}\cdot\boldsymbol{\Lambda}^{(\eta)}\right)\right)\right\} \nonumber \\
 & + & 3\Bigl\{
 \left(e_{i}\Lambda_{j}^{(\rho)}+e_{j}\Lambda_{i}^{(\rho)}\right)f_{6}^{(a)} \nonumber\\
&+&\left(e_{i}\Lambda_{j}^{(\eta)}+e_{j}\Lambda_{i}^{(\eta)}\right)f_{8}^{(a)}\Bigr\} .\nonumber 
\end{eqnarray}
where $\mathbf{\boldsymbol{\Lambda}}_i^{(\rho)}=\boldsymbol{\nabla}_i\log\overline{\rho}$
are components of the gradient of the mean density, the $\mathbf{\boldsymbol{\Lambda}}^{(\eta)}=C_{v}\boldsymbol{\nabla}\log\left(\eta_{T}\right)$
is the same for the turbulent diffusivity. The free parameter $C_{v}$
influences the distribution of the kinetic $\alpha$-effect near the
bottom of the convection zone and the strength of the diamagnetic
pumping there (see below). The $\alpha_{ij}^{(M)}$ reads: 
\begin{equation}
\alpha_{ij}^{(M)}=2f_{2}^{(a)}\delta_{ij}-2f_{1}^{(a)}e_{i}e_{j},\label{alpM}
\end{equation}

The helicity density of the fluctuating part of magnetic field, $\overline{\chi}=\overline{\mathbf{a}\cdot\mathbf{b}}$,
is governed by the conservation law: 
\begin{equation}
\frac{\partial\overline{\chi}^{(tot)}}{\partial t}=-\frac{\overline{\chi}}{R_{m}\tau_{c}}-2\eta\overline{\mathbf{B}}\cdot\mathbf{\overline{J}}-\boldsymbol{\nabla\cdot}\boldsymbol{\boldsymbol{\mathcal{F}}}^{\chi},\label{eq:helcon-1}
\end{equation}
where $\overline{\chi}^{(tot)}=\overline{\chi}+\overline{\mathbf{A}}\cdot\overline{\mathbf{B}}$
is the total magnetic helicity density and
 the $\boldsymbol{\boldsymbol{\mathcal{F}}}^{\chi}=-\eta_{\chi}\boldsymbol{\nabla}\overline{\chi}$
is the diffusive flux of the magnetic helicity, with the $\eta_{\chi}=\frac{3}{10}\left(2f_{1}^{(a)}-f_{2}^{(d)}\right)\eta_{T}$,
$f_{1,2}^{(a)}$ are the functions of the Coriolis number, $R_{m}$
is the magnetic Reynolds number. The $\eta_{\chi}$ is factor ten
smaller than the isotropic part of the magnetic diffusivity \citep{mitra10}.

The quenching functions  which define dependence
of the $\alpha$-effect on the Coriolis number are 
\begin{eqnarray*}
f_{2}^{(a)} & = & \frac{1}{4\Omega^{*\,2}}\left(\left(\Omega^{*\,2}+1\right)\frac{\arctan\Omega^{*}}{\Omega^{*}}-1\right),\\
f_{3}^{(a)} & = & \frac{1}{4\Omega^{*\,2}}\left(2-2\frac{\arctan\Omega^{*}}{\Omega^{*}}\right),\\
f_{4}^{(a)} & = &
\frac{1}{6\Omega^{*\,3}}\Bigl(3\left(\Omega^{*4}+6\Omega^{*2}+5\right)\frac{\arctan\Omega^{*}}{\Omega^{*}}\\
&-& \left(13\Omega^{*2}+15\right)\Bigr),\\
f_{5}^{(a)} & = & \frac{1}{\Omega^{*}}\left(\left(\Omega^{*2}+3\right)\frac{\arctan\Omega^{*}}{\Omega^{*}}-3\right),\\
f_{6}^{(a)} & = & \frac{1}{6\Omega^{*\,3}}\left(3\left(\Omega^{*2}+2\right)\frac{\arctan\Omega^{*}}{\Omega^{*}}-\left(\Omega^{*2}+6\right)\right),\\
f_{8}^{(a)} & = & -\frac{1}{12\Omega^{*\,3}}\left(3\left(4\Omega^{*2}+2\right)\frac{\arctan\Omega^{*}}{\Omega^{*}}\right.\\
 & - & \left.\left(10\Omega^{*2}+6\right)\right),\nonumber \\
f_{10}^{(a)} & = & \frac{1}{\Omega^{*}}\left(1-\left(\Omega^{*2}+1\right)\frac{\arctan\Omega^{*}}{\Omega^{*}}\right),\\
f_{11}^{(a)} & = & -\frac{1}{6\Omega^{*\,3}}\left(3\left(\Omega^{*2}+1\right)^{2}\frac{\arctan\Omega^{*}}{\Omega^{*}}-\left(5\Omega^{*2}+3\right)\right),
\end{eqnarray*}
Functions $f^{(a)}_{2-11}$ were defined in P08 for the general case which
includes the effects the hydrodynamic and magnetic fluctuations in
the background turbulence. Here, we write the expressions for the
case when the background turbulent fluctuations of the small-scale
magnetic field are in equipartition with the hydrodynamic fluctuations,
i.e., ${\displaystyle \varepsilon=\frac{b'^{2}}{4\pi\overline{\rho}u'^{2}}}=1$,
where the $u'^{2}$ and $b'^{2}$ are intensity of the background
turbulent velocity and magnetic field. The magnetic quenching function
of the hydrodynamic part of $\alpha$-effect is as follows
\begin{equation}
\psi_{\alpha}=\frac{5}{128\beta^{4}}\left(16\beta^{2}-3-3\left(4\beta^{2}-1\right)\frac{\arctan\left(2\beta\right)}{2\beta}\right).\nonumber
\end{equation}
Note in the notation of P08 $\psi_{\alpha}=-3/4\phi_{6}^{(a)}$. 

\subsubsection{The turbulent pumping}
The turbulent pumping of the mean-field is the sum of the contributions
due to the mean density gradient \citep{kit:1991}, $\gamma_{ij}^{(\rho)}$,
the diamagnetic pumping \citep{1987SvAL...13..338K}, $\gamma_{ij}^{(\eta)}$,
the mean-field magnetic buoyancy \citep{1993A&A...274..647K}, $\gamma_{ij}^{(b)}$,
and effects  due to the large-scale shear \citep{garr2011}, $\gamma_{ij}^{(H)}$: 
\begin{equation}
\gamma_{ij}=\gamma_{ij}^{(\rho)}+\gamma_{ij}^{(\eta)}+\gamma_{ij}^{(b)}+\gamma_{ij}^{(H)},\label{eq:pump}
\end{equation}
where each contribution is defined as follows:
\begin{eqnarray}
\gamma_{ij}^{(\rho)} & = & 3\eta_{T}\left\{ f_{3}^{(a)}\Lambda_{n}^{(\rho)}+f_{1}^{(a)}\left(\mathbf{e}\cdot\boldsymbol{\Lambda}^{(\rho)}\right)e_{n}\right\} \varepsilon_{inj}\label{eq:pumpr}\\
 &  & -3\eta_{T}f_{1}^{(a)}e_{j}\varepsilon_{inm}e_{n}\Lambda_{m}^{(\rho)}\nonumber \\
\gamma_{ij}^{(\eta)} & = & \!\!\frac{3}{2}\eta_{T}\left\{ f_{2}^{(a)}\Lambda_{n}^{(\eta)}\varepsilon_{inj}\!\!+\!\!f_{1}^{(a)}e_{j}\varepsilon_{inm}e_{n}\Lambda_{m}^{(\eta)}\varepsilon_{inm}\right\} \label{eq:pumpu}\\
\gamma_{ij}^{(b)} & = & \frac{\alpha_{MLT}u'}{\gamma}\beta^{2}K\left(\beta\right)g_{n}\varepsilon_{inj},\label{eq:pumpb}\\
\gamma_{ij}^{(H)} & = & \left(f_{2}^{(\gamma)}\frac{\overline{\chi}}{\overline{\rho}\ell^{2}}+f_{1}^{(\gamma)}h_{\mathcal{K}}\right)\tau_{c}^{2}\varepsilon_{ikj}\overline{W}_{k}\label{eq:pumpH}
\end{eqnarray}
where $f_{1,2,3}^{(a)}\left(\Omega^{\star}\right)$ and $f_{1,2}^{(\gamma)}\left(\Omega^{\star}\right)$
- are functions of the Coriolis number, $u'$ is the RMS of the convective
velocity, the $\alpha_{MLT}=2$ is the parameter of the mixing length
theory (the model by Stix 2002),
$\gamma=5/3$ is the adiabatic exponent, $\mathbf{g}$ is the
unit vector in the radial direction. The quenching of the magnetic buoyancy (see \citep{1993A&A...274..647K})
is determined by
\[
K=\frac{1}{16\beta^{4}}\left(\frac{\left(\beta^{2}-3\right)}{\beta}\arctan\left(\beta\right)+\frac{\left(\beta^{2}+3\right)}{\left(\beta^{2}+1\right)}\right).\]
f{It is noteworthy
that the mean-field magnetic buoyancy produces the mean drift of the
horizontal magnetic field  toward the surface.} This 
 reduces the efficiency of the dynamo process in the deep convection
zone. The drift velocity is proportional to the pressure of the large-scale
magnetic field. The $K(\beta)$ takes into
account the feedback of the
magnetic tensions on the buoyancy effect.  For the magnetic field with
strength  of about the equipartition
value, $\beta\sim1$, the drift velocity is a few meters per second. 
The pumping effect $\gamma_{ij}^{(H)}$
is governed by the helicity parameters including the magnetic, $\overline{\chi}$,
and kinetic helicity, $h_{\mathcal{K}}={\displaystyle \frac{3\eta_{T}}{2\tau_{c}}F_{1}\left(\Omega^{*}\right)\cos\theta\frac{\partial}{\partial r}}\log\left(\bar{\rho}u'\right)$
(see, \citealt{kps:06}), and the large-scale vorticity vector $\overline{\mathbf{W}}=\mathbf{\nabla}\times\overline{\mathbf{U}}$.
Dependence of the $\gamma_{ij}^{(H)}$ on the Coriolis number is controlled
by the quenching functions:
\begin{eqnarray*}
f_{1}^{(\gamma)} & = & \frac{1}{(24\Omega^{\ast})^{2}}\left(\left(1300\Omega^{\ast2}+391\right)\frac{\arctan\left(2\Omega^{\ast}\right)}{2\Omega^{\ast}}\right.\\
 & - & \left.1456\left(\Omega^{\ast2}+1\right)\frac{\arctan\left(\Omega^{\ast}\right)}{\Omega^{\ast}}-3\left(32\Omega^{\ast2}-355\right)\right),\nonumber \\
F_{1} & = & \frac{1}{2\Omega^{*}}\left(\left(\Omega^{*2}-1\right)\frac{\arctan\Omega^{*}}{\Omega^{*}}+1\right)
\end{eqnarray*}
and $f_{2}^{\left(\gamma\right)}=3f_{2}^{\left(a\right)}$. 
 \citet{2013GApFD.107..185P} analyzed the total effect
of the $\gamma_{ij}$ in details (see Fig.4 there). It was found that
for the solar case, the $\gamma_{ij}$ results to the downward pumping
in the most part of the convection zone. The velocity of the drift
is a few meter per second in the main part of the convection zone.
The magnitude of the latitudinal pumping of the toroidal magnetic
field to equator is about 1 m/s.

\subsection{The turbulent stresses}

The turbulent stresses take into account the turbulent viscosity and
generation of the large-scale shear due to the $\Lambda$- effect
\citep{1999A&A...344..911K}: 
\begin{eqnarray}
T_{r\phi} & = & \overline{\rho}\nu_{T}\left\{ \Phi_{\perp}+\left(\Phi_{\|}-\Phi_{\perp}\right)\mu^{2}\right\} r\frac{\partial\sin\theta\Omega}{\partial r}\nonumber \\
 & + & \overline{\rho}\nu_{T}\sin\theta\left(\Phi_{\|}-\Phi_{\perp}\right)\left(1-\mu^{2}\right)\frac{\partial\Omega}{\partial\mu}\label{eq:trf}\\
 & - & \overline{\rho}\nu_{T}\sin\theta\Omega\left(\frac{\alpha_{MLT}}{\gamma}\right)^{2}\left(V^{(0)}+\sin^{2}\theta V^{(1)}\right),\nonumber \\
T_{\theta\phi} & = & \overline{\rho}\nu_{T}\sin^{2}\theta\left\{ \Phi_{\perp}+\left(\Phi_{\|}-\Phi_{\perp}\right)\sin^{2}\theta\right\} \frac{\partial\Omega}{\partial\mu}\nonumber \\
 & + & \overline{\rho}\nu_{T}\left(\Phi_{\|}-\Phi_{\perp}\right)\mu\sin^{2}\theta r\frac{\partial\Omega}{\partial r}\label{eq:ttf}\\
 & + & \overline{\rho}\nu_{T}\mu\Omega\sin^{2}\theta\left(\frac{\alpha_{MLT}}{\gamma}\right)^{2}\left(H^{(0)}+\sin^{2}\theta H^{(1)}\right),\nonumber 
\end{eqnarray}
where $\mu=\cos\theta$, $\nu_{T}={\displaystyle \frac{4}{5}\eta_{T}}$. The viscosity
functions - $\Phi_{\|},\Phi_{\perp}$ and the $\Lambda$- effect -
$V^{\left(0,1\right)}$ and $H^{\left(0,1\right)}$, depend
on the Coriolis number and the strength of the large-scale magnetic
field. They also depend on the anisotropy of the convective flows.
In following to \citet{1999SoPh..189..227K,kit2004AR} and \citet{phd}
we employ the following expressions: 
\begin{eqnarray}
\Phi_{\perp}\left(\Omega^{\star},\beta\right)\!\!\!\!&\!=\!&\!\!\psi_{\perp}\left(\Omega^{\star}\right)\left(\Phi_{q}\psi_{1}\left(\beta\right)
\!+\!\left(1\!\!-\!\!\Phi_{q}\right)\phi_{V\perp}\left(\beta\right)\right),\\
\Phi_{\parallel}\left(\Omega^{\star},\beta\right) & = & \psi_{\parallel}\left(\Omega^{\star}\right)\left(\Phi_{q}\psi_{1}\left(\beta\right)+\left(1-\Phi_{q}\right)\phi_{V}\left(\beta\right)\right),\\
V^{(0)} & = & \left\{ J_{0}\left(\Omega^{\star}\right)+J_{1}\left(\Omega^{\star}\right)+a\left(I_{0}\left(\Omega^{\star}\right)+I_{1}\left(\Omega^{\star}\right)\right)\right\} \nonumber \\
 & \times & \left(\Phi_{q}K_{1}\left(\beta\right)+\left(1-\Phi_{q}\right)\phi_{V}\left(\beta\right)\right),\\
V^{(1)} & = & \left\{ J_{1}\left(\Omega^{\star}\right)+aI_{1}\left(\Omega^{\star}\right)\right\} \nonumber \\
 & \times & \left(\Phi_{q}K_{1}\left(\beta\right)+\left(1-\Phi_{q}\right)\phi_{V}\left(\beta\right)\right),\\
H^{(0)} & = & J_{4}\left(\Omega^{\star}\right)\phi_{H}\left(\beta\right),\\
H^{(1)} & = & -V^{(1)},
\end{eqnarray}
where, ${\displaystyle {\Phi_{q}=\frac{\arctan\Omega^{\star}}{\Omega^{\star}}}}$.
We employ the quenching functions which were derived in the previous
studies, e.g., the $\psi_{\perp,\parallel}$, $\psi_{1}$, $K_{1}$
and $J_{0,1}$ can be found in \citep{1993A&A...279L...1K,1994AN....315..157K,kuetal96},
the $I_{0,1}$ - in \citep{kit2004AR}. For convenience we put them
here, 
\begin{eqnarray}
\psi_{\parallel} & = &
\frac{15}{32\Omega^{*4}}\Bigl(21-3\Omega^{*2}+\frac{4\Omega^{*2}}{1+\Omega^{*2}} \\
&-& \left(21+4\Omega^{*2}-\Omega^{*4}\right)\frac{\arctan\Omega^{*}}{\Omega^{*}}\Bigr),\nonumber\\
\psi_{\perp} & = &
\frac{15}{128\Omega^{*4}}\Bigl(\Omega^{*2}-21-\frac{8\Omega^{*2}}{1+\Omega^{*2}} \\
&+&\left(21+14\Omega^{*2}+\Omega^{*4}\right)\frac{\arctan\Omega^{*}}{\Omega^{*}}\Bigr).
\end{eqnarray}
\begin{eqnarray}
J_{0} & = & \frac{1}{2\Omega^{*4}}\left(9-\frac{2\Omega^{*2}}{1+\Omega^{*2}}-\left(\Omega^{*2}+9\right)\frac{\arctan\Omega^{*}}{\Omega^{*}}\right),\\
J_{1} & = &
\frac{1}{2\Omega^{*4}}\Bigl(\frac{45+42\Omega^{*2}+\Omega^{*4}}{1+\Omega^{*2}}\\
&+&\left(\Omega^{*4}-12\Omega^{*2}-459\right)\frac{\arctan\Omega^{*}}{\Omega^{*}}\Bigr),\nonumber\\
I_{0} & = & -\frac{3}{4\Omega^{*4}}\left(\frac{3+\Omega^{*2}}{1+\Omega^{*2}}+\left(\Omega^{*2}-3\right)\frac{\arctan\Omega^{*}}{\Omega^{*}}\right),\\
I_{1} & = &\frac{3}{4\Omega^{*4}}\Bigl(\frac{2\Omega^{*2}}{1+\Omega^{*2}}-15 \\
&+&\left(3\Omega^{*2}+15\right)\frac{\arctan\Omega^{*}}{\Omega^{*}}\Bigr),\nonumber\\
K_{1} & = & \frac{1}{16\beta^{4}}\left(\frac{\beta^{2}+1}{\beta}\arctan\beta-1-\frac{2\beta^{2}}{3\left(1+\beta^{2}\right)}\right)
\end{eqnarray}
The magnetic quenching functions $\phi_{H,V}$ follows from the general
expression for the $\Lambda$ - effect and the turbulent viscosity
for the fast rotating regime ($\Omega^{\star}>1$) given by \citet{kuetal96}
and by \citet{phd}. They are defined as follows: 
\begin{eqnarray}
\phi_{V\perp} & = & \frac{4}{\beta^{4}\sqrt{\left(1+\beta^{2}\right)^{3}}}\left(\left(\beta^{4}+19\beta^{2}+18\right)\sqrt{\left(1+\beta^{2}\right)}\right.\nonumber \\
 & - & \left.8\beta^{4}-28\beta^{2}-18\right),\\
\phi_{V} & = & \frac{2}{\beta^{2}}\left(1-\frac{1}{\sqrt{\left(1+\beta^{2}\right)}}\right),\\
\phi_{H} & = & \frac{4}{\beta^{2}}\left(\frac{2+3\beta^{2}}{2\sqrt{\left(1+\beta^{2}\right)^{3}}}-1\right).
\end{eqnarray}
The Eqs.(A13-A30) take into account the different regimes of the magnetic
quenching of the $\Lambda$-effect and the eddy viscosity in the cases
of the slow ($\Omega^{\star}\ll1$) and the fast rotation ($\Omega^{\star}\gg1$).
In particular, for the case $\Omega^{\star}\ll1$ the turbulent viscosity
is quenched as $\beta^{-1}$ when $\beta>1$ , and for the case $\Omega^{\star}\gg1$
there is a stronger quenching, $\beta^{-2}$. Thus, in the presence of
the strong large-scale
toroidal magnetic field, the ratio between the eddy viscosity along and
perpendicular to  rotation axis is decreased \citep{phd}.
\end{document}